\documentclass[hyper,11pt,paper]{article}
\usepackage{jheppub}
\usepackage[usenames,dvipsnames,table]{xcolor}
\usepackage{graphicx,amsmath,amssymb,multirow,array,bm,mathrsfs}
\usepackage{epsf,amsfonts}
\usepackage[numbers,sort&compress]{natbib}
\DeclareMathOperator\arctanh{arctanh}
\title{
An anomalous propulsion mechanism
}
\author{Evgeny Shaverin and Amos Yarom}

\affiliation{Department of Physics, Technion, Haifa 32000, Israel}

\emailAdd{evgeny@tx.technion.ac.il, ayarom@physics.technion.ac.il}
\abstract{
We consider a gas of free chiral fermions trapped inside a uniform rotating spherical shell. Once the shell becomes transparent the fermions are emitted along the axis of rotation due to the chiral and mixed anomaly. In return, owing to momentum conservation, the shell is propelled forward. We study the dependence of the magnitude of this effect on the shell parameters in a  controlled setting and find that it is sensitive to the formation of an ergosphere around the rotating shell. A brief discussion on a possible relation to pulsar kicks is provided.
}
\begin{document}
\maketitle

\section{Introduction }
\label{S:intro}

Chiral fermions are anomalous---what seems like a $U(1)$ symmetry of the free action is not a symmetry of the partition function. This peculiar property has had many repercussions on modern-day particle physics ranging from an explanation of the abnormally large decay rate of the neutral pion \cite{Adler:1969gk,Bell:1969ts} through restrictions on possible extensions of the standard model \cite{hooft1980recent} to duality relations between supersymmetric quantum field theories \cite{Seiberg:1994bz}. Over the last forty odd years much effort has gone into a comprehensive classification of anomalies and an understanding of their effect on S-matrix elements and vacuum correlation functions (see e.g., \cite{bertlmann2000anomalies} for a comprehensive review). More recently, it has been established that anomalies play a  prominent role in the response of the system to vorticity (or a magnetic field) when the system is in or near thermal equilibrium \cite{Erdmenger:2008rm,Banerjee:2008th,Son:2009tf,Neiman:2010zi,Landsteiner:2011cp,Jensen:2012kj,Jensen:2013kka,Jensen:2013rga}. 

Let us briefly review these latest developments. The (covariant) energy momentum tensor $T^{\mu\nu}$ and chiral current $J^{\mu}$ satisfy the (non-)conservation laws
\begin{align}
\begin{split}
\label{E:conservationlaws} 
	\nabla_{\mu}J^{\mu}  &= \frac{1}{4} \epsilon^{\mu\nu\rho\sigma} \left( 3 c_A F_{\mu\nu}F_{\rho\sigma} + c_m R^{\alpha}{}_{\beta\mu\nu} R^{\beta}{}_{\alpha\rho\sigma} \right)\\
	\nabla_{\mu}T^{\mu\nu}&= F^{\mu}{}_{\nu} J^{\nu} + \frac{c_m}{2} \nabla_{\nu} \left( \epsilon^{\rho\sigma\alpha\beta} F_{\rho\sigma}R^{\mu\nu}{}_{\alpha\beta} \right) 
\end{split}
\end{align} 
where
\begin{equation}
	c_A = -\frac{1}{24\pi^2}
	\qquad
	c_m= -\frac{1}{192 \pi^2}\,,
\end{equation}
are the anomaly coefficients for a left handed fermion and indicate the strength of the anomaly, $R^{\alpha}{}_{\beta\gamma\delta}$ is the Riemann tensor and $F=dA$ is a flavor field strength. Our conventions are the same as those specified in \cite{Jensen:2013kka}, where \eqref{E:conservationlaws} has been carefully derived. Had $c_A=c_m=0$, \eqref{E:conservationlaws} would have reduced to current conservation and energy-momentum conservation up to a Joule heating term.

When the system is near thermodynamic equilibrium one finds that the 10 components of the stress tensor and four components of the chiral current depend on five parameters: a temperature field $T$, a chemical potential field $\mu$, and a velocity field $u^{\mu}$ normalized such that $u^{\mu}u_{\mu} = -1$. The expression relating the components of the stress tensor and current to the five thermodynamic parameters are referred to as constitutive relations. Due to possible field redefinitions of the thermodynamic parameters there is some non physical ambiguity in the constitutive relations which may be fixed by an appropriate definition of thermodynamic fields (see e.g., \cite{landau1987course,Bhattacharya:2011eea,Bhattacharya:2011tra}). Choosing a particular definition is referred to as a choice of frame. In what follows we will use a particular definition, referred to as the thermodynamic frame \cite{Jensen:2012jh} (used also in \cite{Landsteiner:2012kd}). We point out that our final results are frame independent. 

The constitutive relations for the energy momentum tensor, to first order in derivatives, in the thermodynamic frame, are given by
\begin{subequations}
\label{E:constitutive}
\begin{equation}
	T^{\mu\nu} =  \epsilon u^{\mu} u^{\nu} + P (g^{\mu\nu} + u^{\mu} u^{\nu}) + u^{\mu} q^{\nu} + u^{\nu} q^{\mu}  - \eta \sigma^{\mu\nu}
\end{equation}
where $P(T,\mu)$ is the pressure, $\epsilon = -P + T \frac{\partial P}{\partial T} + \mu \frac{\partial P}{\partial \mu}$, 
\begin{equation}
\label{E:qconstitutive}
	q^{\mu} = -2 \left( 8 \pi^2 c_m \mu T^2 + c_A \mu^3\right) \epsilon^{\mu\nu\rho\sigma}u_{\nu} \partial_{\rho} u_{\sigma}
\end{equation}
and
\begin{equation}
	\sigma^{\mu\nu} = (u^{\mu}u^{\rho} + g^{\mu\rho}) (u^{\nu}u^{\sigma} + g^{\nu\sigma}) \left( \nabla_{\rho} u_{\sigma} + \nabla_{\sigma} u_{\rho}\right) - \frac{1}{3}\left(u^{\mu}u^{\nu} + g^{\mu\nu} \right) \nabla_{\rho}u^{\rho}\,.
\end{equation}
The constitutive relations for the anomalous $U(1)$ current are given by
\begin{equation}
	J^{\mu} = \rho u^{\mu}+\nu^{\mu} \,.
\end{equation}
where $ \rho $ is a charge density and $ \nu^{\mu} $ is given by 
\begin{equation}
\label{E:nuconstitutive}
	\nu^{\mu} = \sigma  (u^{\mu}u^{\rho} + g^{\mu\rho}) \left(E_{\rho} - T \nabla_{\rho} \frac{\mu}{T}  \right) -(8 \pi^2 c_m T^2+3 c_A \mu^2)\epsilon^{\mu\nu\rho\sigma}u_{\nu} \partial_{\rho} u_{\sigma} \, 
\end{equation}
\end{subequations}
where $E^{\mu} = F^{\mu\nu}u_{\nu}$ is the flavor electric field in the rest frame of a fluid element.
We refer the reader to \cite{Jensen:2012kj} for a detailed derivation of \eqref{E:constitutive}.
Solving \eqref{E:conservationlaws} with \eqref{E:constitutive} will provide us with an expression for the velocity field $u^{\mu}$, chemical potential $\mu$ and temperature $T$ in terms of the background metric and gauge field and the boundary and initial data. 

In section \ref{S:rotatingfermions} we will present a solution to \eqref{E:conservationlaws} for a chiral fermion gas trapped inside a uniformly rotating spherical shell. However, before going into the details of such a computation we point out that \eqref{E:qconstitutive} and \eqref{E:nuconstitutive} with non zero $c_A$ or $c_m$ indicate that the chiral current and heat current respond in an unexpected way to the vorticity vector, $\omega^{\mu} = \epsilon^{\mu\nu\rho\sigma} u_{\nu}\partial_{\rho} u_{\sigma}$. If the gas is forced to rotate then, at the level of linear response, a current will be generated in the direction of rotation.

It is somewhat surprising that the anomaly modifies the hydrodynamic behavior of a macroscopic system. The unusual dependence of the current on vorticity holds the promise of observing the effect of anomalies in appropriate astrophysical or cosmological settings. Our initial foray in this direction involves a study of the behavior of a bound rotating gas of chiral fermions which is suddenly free to expand. We will argue below that due to the anomaly the fermions will be emitted in the direction of rotation and, owing to momentum conservation, will propel the vessel which bound them in the opposite direction. 

It is tempting to relate the aforementioned propulsion mechanism to an astrophysical phenomenon referred to as pulsar kicks whereby young neutron stars move at high linear velocities of order of hundreds of kilometers per second. These peculiarly high velocities are much larger than the velocity of the progenitor star which core-collapsed, exploded and generated the neutron star and are difficult to explain by conventional means, see e.g., \cite{Lai:1999za}. 

The idea that anomalies are responsible for pulsar kicks has appeared sporadically in the literature, see, e.g., \cite{Kusenko:1998yy,Sagert:2007ug,Charbonneau:2009ax,Charbonneau:2010jx,Kaminski:2014jda}. Often, the kick is attributed to the response of an anomalous current to a magnetic field (which is frequently referred to as the chiral magnetic effect \cite{Kharzeev:2009fn}), the reason being that the response of the anomalous current to vorticity is negligible. Indeed, as we will see below, if the geometry exterior to the shell does not posses an ergosphere then the angular velocity of the shell is too small to be observable. However, when the geometry outside the shell possesses an ergosphere (or if the radius of the sphere is close to its Schwarzschild radius) then standard arguments associated with dimensional analysis are somewhat incomplete.

Our work is organized as follows. In section \ref{S:rotatingfermions} we solve \eqref{E:conservationlaws} for a gas of fermions which is trapped in a uniformly rotating, compact, axisymmetric body. We compute the momentum of the emitted fermions once the body becomes transparent under the assumption that the emission is instantaneous. An analysis of the result and its relation to pulsar kicks can be found in section \ref{S:discussion}.

\section{Rotating chiral fermions}
\label{S:rotatingfermions}
Consider an axisymmetric body of mass $M$ and radius $R$ rotating about a fixed axis with angular velocity $\omega$ as measured by an observer at flat asymptotic infinity. 
If we parameterize the metric at infinity via a standard spherical coordinate system,
\begin{equation}
	ds^2 = -dt^2 + dr^2 + r^2 \left(d\theta^2 + \sin^2\theta d\phi^2\right)
\end{equation}
then the rotational symmetry of the problem amounts to the existence of Killing vectors $\partial_t$ and $\partial_\phi$ and two $\mathbb{Z}_2$ symmetries: one which flips time and the direction of rotation ($t\to-t$ and $\phi \to-\phi$), and another which reflects across the equator, ($\theta \to \pi-\theta$). Given this symmetry we may choose the $r$ and $\theta$ coordinates such that the metric takes the form
\begin{equation}
	ds^2 = g_{ij}(r,\theta) dx^i dx^j + g_{rr}(r,\theta)(dr^2 + r^2 d\theta^2)
\end{equation}
with Roman indices $i$ and $j$ running over $\phi$ and $t$. 
Often we will switch to a cylindrical coordinate system 
\begin{equation}
	ds^2 = g_{ij}(\rho,z) dx^i dx^j + g_{zz}(\rho,z)(d\rho^2 + dz^2)
\end{equation}
where $\rho = r \sin\theta$ and $z= r\cos\theta$.


Let us assume that the mass $M$ contains a thermally equilibrated gas of chiral fermions whose energy is small enough compared to $M$ so that they don't modify the space-time metric. We also assume that modifications to the energy momentum tensor of the mass due its interaction with the fermions may be ignored. Put differently, we assume that the fermions are bound to the interior of the rotating body via its gravitational pull, and neglect the behavior of the fermion gas at the boundary of the body. 

Since the metric due to the rotating body is stationary we may use the techniques developed in \cite{Banerjee:2012iz,Jensen:2012jh} to compute the stress tensor and current of the fermion gas in such a background. In the thermodynamic frame, the solution is guaranteed to be given by
\begin{equation}
\label{E:solution}
	u^{\mu} = \frac{1}{\sqrt{-g_{tt}}} \begin{pmatrix} 1 \\ 0 \\ 0 \\ 0 \end{pmatrix} 
	\qquad
	T = \frac{T_0}{\sqrt{-g_{tt}}}
	\qquad
	\frac{\mu}{T} =\frac{A_t}{T_0}\,.
\end{equation}
Indeed, one can check that upon insertion of \eqref{E:solution} into \eqref{E:constitutive} the conservation equations \eqref{E:conservationlaws} are satisfied. The reader is referred to \cite{Banerjee:2012iz,Jensen:2012jh,Jensen:2013kka} for details.

At this point, it is worth pausing to emphasize that the final expression obtained for the stress tensor after inserting \eqref{E:solution} into \eqref{E:constitutive} is independent of the choice of frame. Had we used a different set of constitutive relations, the expressions \eqref{E:solution} would have been modified such that the resulting dependence of the energy momentum tensor and current on the background fields would have been the same as the one obtained from \eqref{E:solution} and \eqref{E:constitutive}. We have chosen to present the thermodynamic frame in \eqref{E:constitutive} since the techniques of \cite{Banerjee:2012iz,Jensen:2012jh} provide us with a nifty solution to the equations of motion in such a frame. Had we worked in the more traditional Landau frame our final result for the stress tensor would have remained unchanged.

With the solution \eqref{E:solution} at hand, one can now compute the four-momentum of the fermion gas
\begin{equation}
	P^{\mu} = \int T^{\mu0} \sqrt{g_{\phi\phi}}g_{zz} d\rho  dz d\phi
\end{equation}
where the integral is over the entire volume encasing the gas. Of particular interest is the pressure along the axis of rotation, $P^z$. A short computation yields
\begin{align}
\label{E:E}
	P^0 &= 2\pi \int \frac{3 g_{tt}g_{\phi\phi} - 4 g_{t\phi}^2}{-g_{tt}(g_{tt}g_{\phi\phi} - g_{t\phi}^2)} P(T,\mu)  g_{rr} \sqrt{g_{\phi\phi}}dr dz \\
\label{E:Pz}
	P^z &=-2\pi \int 2 \mu \sqrt{g_{\phi\phi}} ( 8\pi^2 c_m T^2 + c_A \mu^2) \frac{g_{t\phi}\partial_\rho g_{tt} - g_{tt} \partial_\rho g_{t\phi} }{(-g_{tt})^{3/2} \sqrt{g_{t\phi}^2 - g_{tt} g_{\phi\phi}}} dr dz
\end{align}
with 
\begin{align}
\begin{split}
	P(T,\mu) & = \frac{1}{3} \int \frac{d^3 p}{(2\pi)^3} \, p \left[\frac{1}{e^{(p-\mu)/T}} + \frac{1}{e^{(p+\mu)/T}}\right] \\
	& = \frac{7 \pi^2 T^4}{360}+\frac{T^2 \mu^2}{12}+\frac{\mu^4}{24 \pi^2}
\end{split}
\end{align}
the pressure of a chiral fermion gas at temperature $T$ and chemical potential $\mu$. We have not written down similar expressions for the remaining components of the momentum four-vector of the gas.

Let us assume that the rotating body becomes suddenly transparent to the fermion gas. The velocity gained by the rotating body due to the ejection of the gas is given by
\begin{equation}
\label{E:main}
	v = -\frac{P^z}{M} \,.
\end{equation}

\section{Discussion}
\label{S:discussion}
To understand the implications of our main result \eqref{E:main} it is convenient to consider a particular configuration where the components of the space-time metric are available explicitly. Unfortunately, a full analytic solution for a rotating axisymmetric body is not known in general. Nevertheless, for slowly rotating shells such a result is available when working perturbatively in the shell angular velocity. In appendix \ref{A:shellmetric} we show that
\begin{subequations}
\label{E:slowrotation}
\begin{align}
\begin{split}
\label{E:slowmetric}
	g_{tt} =& -\frac{(R-r_s)^2}{(R+r_s)^2} \left(1 + \mathcal{O}(\omega^2) \right) + \omega^2 \frac{R^2 \rho^2 \lambda^2}{(R+r_s)^8}\left(1 + \mathcal{O}(\omega^4) \right)\\
	g_{t\phi} =& - \frac{\rho^2 \lambda_0 \omega}{R(R+r_s)^2}\left(1 + \mathcal{O}(\omega^2)\right) \\
	g_{\phi\phi} =& \frac{\rho^2 (R+r_s)^4}{R^4}\left(1 + \mathcal{O}(\omega^2)\right)  \\
	g_{rr} = & \frac{(R+r_s)^4}{R^4} \left(1 + \mathcal{O}(\omega^2)\right)  \\
	g_{zz} = & g_{rr}
\end{split}
\end{align}
with 
\begin{equation}
	\lambda_0 = \frac{4 (2 R-r_s) r_s (R+r_s)^5}{R^3 (3 R-r_s)}\,,
\end{equation}
\end{subequations}
solves the Einstein equations for a uniformly rotating thin spherical shell. Here $r_s$ is the Schwarzschild radius. In our coordinate system it is related to the mass of the shell via $r_s =M/2$. In the first line of \eqref{E:slowmetric} we have kept the explicit $\mathcal{O}(\omega^2)$ dependence in order to be able to account for the case of $R \sim r_s$. The result \eqref{E:slowmetric}, not including the $\mathcal{O}(\omega^2)$ corrections has been first obtained in \cite{brill1966rotating} and is most relevant for the ensuing discussion.

In order to understand the role of the various metric components in \eqref{E:Pz} let us insert the solution \eqref{E:slowrotation} into \eqref{E:E} and \eqref{E:Pz} together with the assumption that the chemical potential is constant throughout and that we can neglect the temperature relative to the chemical potential, $\mu/T \gg 1$.\footnote{As we will see shortly, we will be particularly interested in the case of small $g_{tt}$. As long as $g_{tt}$ is small but finite we may always tune $T_0$ in \eqref{E:solution} so that the chemical potential is larger than the temperature.}
We find
\begin{subequations}
\begin{align}
\label{E:PzP0vals}
	P^z =& -\frac{16 \pi c_A (R+r_s)^6\mu^3}{R^2 \lambda_0\omega} \left(1-\frac{(R-r_s)(R+r_s)^3}{R^2 \lambda_0 \omega} \arctanh\left(\frac{R^2 \lambda_0\omega}{(R-r_s)(R+r_s)^3}\right)\right) \\
\intertext{and}
\label{E:P0eq}
	P^0= & \frac{ \mu^4 (R+r_s)^{14}}{2\pi R^7 \lambda_0^2 \omega^2} \left( 1 - \frac{1}{\Xi} \arctan \Xi \right) 
\end{align}
\end{subequations}
with
\begin{equation}
	\Xi = \frac{R^2 \lambda_0 \omega}{\sqrt{(R-r_s)^2(R+r_s)^6 - R^4 \lambda_0^2 \omega^2}}\,.
\end{equation}
Strictly speaking \eqref{E:PzP0vals} and \eqref{E:P0eq} should be multiplied by $\left(1+ \mathcal{O}(\omega^2)\right)$.

Note that there exists a critical frequency $\omega_c$,
\begin{equation}
\label{E:Omegacrit}
	\frac{(R-r_s)(R+r_s)^3}{R \lambda_0} =  \omega_c R 
\end{equation}
at which $P^z$ diverges logarithmically. The critical frequency is precisely the value of $\omega$ at which $g_{tt}=0$ and an ergosphere is formed. (We are assuming here that $\omega_c R \ll 1$ so that the approximation \eqref{E:slowmetric} remains valid.)   In what follows we will consider configurations where
\begin{equation}
	\omega = \omega_c (1-\delta)
\end{equation}
with $\delta$ positive. For small values of $\delta$, an ergosphere is just about to form.  If an ergosphere forms outside the spherical shell then the methods of \cite{Banerjee:2012iz,Jensen:2012jh} have to be used with some care in order to keep track of the location of the timelike Killing vector in the background space-time. We leave such an analysis for future research.

Using \eqref{E:PzP0vals} and \eqref{E:main} we find that in the limit where $\omega R \ll 1$ and $e^\delta > 1$, 
\begin{equation}
\label{E:final1}
	\frac{v}{c} = \frac{8}{3} \left(\frac{8}{3\pi}\right)^{1/4}  \frac{\sqrt{1+\xi} (1+2\xi)}{(2+\xi)(2+3\xi)\sqrt{\xi}} \frac{(P^0 R)^{3/4} (\omega R)}{(M R)}
\end{equation}
where we have defined 
\begin{equation}
\label{E:sdef}
	R = r_s(1+\xi)
\end{equation}
and have used \eqref{E:P0eq} to replace the chemical potential with the total energy.
But if $\omega R \ll 1$ and $\delta \ll 1$, then we find
\begin{equation}
\label{E:final2}
	\frac{v}{c} = -\frac{1}{3} \left( \frac{8}{\pi} \right)^{1/4}  \frac{\sqrt{\xi}(1+\xi)^{5/2}}{(2+\xi)^3} \frac{(P^0 R)^{3/4}}{M R} \ln(\delta) \, .
\end{equation}

As mentioned previously, the logarithmic divergence of $v/c$ in \eqref{E:final2} is a result of the formation of an ergosphere around the rotating spherical shell. Due to the general form of \eqref{E:Pz} and \eqref{E:E} it seems possible that the divergence of $P^z$ in such configurations goes beyond our slowly rotating shell model \eqref{E:slowrotation}.

Let us now attempt to relate \eqref{E:final1} or \eqref{E:final2} to pulsar kicks \cite{Lai:1999za}. Recall that once a star does not have enough energy to resist its own gravitational pull, it collapses and forms a proto-neutron star via a supernova explosion. The proto-neutron star evolves into a neutron star via a cooling mechanism. Since both during the collapse process and during the cooling phase, neutrinos are emitted, it is tempting to estimate the recoil of the proto-neutron star to directed neutrino emission via the mechanism we have outlined above. Before doing so we emphasize that our simple-minded model misses many of the important and necessary features of the dynamics of supernova and proto-neutron stars.

Typical values for the mass, $M$, radius, $R_*$, and rotation periods, $T$, of young proto-neutron stars are given by \cite{Goussard:1996dp,Prakash19971}
\begin{equation}
\label{E:NS}
	7 \lesssim R_*/km \lesssim 16\,,
	\qquad
	1 \lesssim M/M_{\odot} \lesssim 2\,, 
	\qquad
	0.4 \lesssim T/\textrm{ms} \lesssim 1.5 \,,
\end{equation}
and the neutrinos are emitted at an estimated \cite{Bethe:1984ux} and measured \cite{Bionta:1987qt,Hirata:1987hu} luminosity of
\begin{equation}
	E_{\nu} \sim 10^{52} \, \textrm{erg s}^{-1} 
\end{equation}
over a period of an order of half a second. 

For typical slowly rotating stars, say, 
\begin{equation}
	R_* \sim 10 \, \textrm{km}\,,
	\qquad
	M \sim 1.5 M_{\odot}\,,
	\qquad
	T \sim 1 \, \textrm{ms} \,,
\end{equation}
we find that
\begin{equation}
	\xi \sim 5.94 \,,
	\qquad
	\omega R \sim 0.16 \,,
	\qquad
	\delta \sim 0.91\,,
\end{equation}	
where we have used
\begin{equation}
	R_*=\int_0^{R} \sqrt{g_{rr}} dr
\end{equation}
with
\begin{equation}
	G = 6.67\times10^{-11} \,\, \textrm{N} \, \textrm{(m/kg)}^2 \, ,
		\qquad
		M_{\odot} = 1.98\times10^{30} \, \textrm{kg}\,.
\end{equation}
The recoil velocity of the shell, according to \eqref{E:final1} comes out to $v \sim 10^{-23} c $, too small to be observable. 
However, if we take more extreme values for the neutron star parameters,
\begin{equation}
	R_* \sim 6.9 \, \textrm{km}\,,
	\qquad
	M \sim 2.2 M_{\odot}\,,
	\qquad
	T \sim 0.5 \, \textrm{ms} \,,
\end{equation}
which imply
\begin{equation}
	\xi \sim 0.68 \,,
	\qquad
	\omega R \sim 0.11 \,,
	\qquad
	\delta \sim 0\,;
\end{equation}	
an ergosphere is generated. In this case, according to \eqref{E:final2}, the ratio $v/c$ may be exceedingly large, it's exact value depending on the precise details of the collapse process which our simple minded model can not capture.  

There are several models which deal with the appearance and instability of an ergosphere in rotating axisymmetric bodies \cite{Friedman,comins1978ergoregion} also in the context of neutron stars \cite{Kokkotas}. It would be interesting to check the viability of our proposed mechanism in a dynamical setting.

\section*{Acknowledgements}
We thank B. Keren-Zur and Y. Oz for useful discussions and sharing some private notes. ES and AY are supported by the ISF under grant numbers 495/11, 630/14 and 1981/14, by the BSF under grant number 2014350, by the European commission FP7, under IRG 908049 and by the GIF under grant number 1156/2011.

\begin{appendix}

\section{The metric of a rotating spherical shell}
\label{A:shellmetric}
The metric induced by a rigidly rotating uniform spherical shell in asymptotically flat space can be computed along the lines of \cite{lewis1932some,Papapetrou,brill1966rotating,DeLaCruz,schutz1978existence,PfisterBraun}. We parameterize our space-time by coordinates $t$, $\rho$, $z$ and $\phi$ with $-\infty<t<\infty$, $\rho>0$, $-\infty<z<\infty$ and $0 \leq \phi<2\pi$. A rigidly rotating sphere will induce an axisymmetric geometry, by which we mean a geometry with two Killing vectors, denoted $\partial_t$ and $\partial_{\phi}$, and two discrete $\mathbb{Z}_2$ symmetries, denoted $T$ and $P$, the first inverts both time, $t$, and the angular direction of rotation, $\phi$ ($t\to-t$ and $\phi \to -\phi$) and the second is a parity symmetry along the equatorial plane ($z \to -z$). The most general line element which satisfies all these symmetries is given by
\begin{equation}
	ds^2 = g_{ij}(\rho,z) dx^i dx^j + g_{ab}(\rho,z) dx^a dx^b
\end{equation}
with Roman indices $i$ and $j$ running over $\phi$ and $t$ and the indices $a$ and $b$ running over $\rho$ and $z$.

One can now argue that there exists a coordinate transformation which diagonalizes $g_{ab}$ leaving us with four independent parameters characterizing the metric. We note in passing that there is still some residual diffeomorphism invariance which allows us to carry out a holomorphic or antiholomorphic coordinate transformation in the $r$, $z$ plane, keeping the diagonal form of $g_{ab}$ fixed. In \cite{lewis1932some} it was shown that, in the absence of matter, one can use these residual coordinate transformations to fix the value of the determinant of $g_{ij}$; in the absence of matter $|g_{ij}|$ satisfies the Laplace equation. In what follows we will not use such a coordinate choice. Instead we will parameterize the line element via
\begin{equation}
\label{E:MyLE}
	ds^2 = -e^{2 U} dt^2 + e^{-2 U}\left( e^{2 K}(dr^2 + r^2 d\theta^2) + W^2(d\phi - \omega A dt)^2 \right)\,,
\end{equation}
similar to \cite{PfisterBraun}. Here $z=r \cos\theta$ and $\rho = r \sin\theta$.

The rigidly rotating shell which sources the metric is assumed to be very thin with delta-function support at $r=R$, $T^{\mu}{}_{\nu} = \tau^{\mu}{}_{\nu} \delta(r-R)$. Since the shell is rigidly rotating then the local velocity field of a shell element, $u^{\mu}$, should satisfy
\begin{equation}
\label{E:energymomentum}
	\tau^{\mu}{}_{\nu} u^{\nu} = - \sigma u^{\nu}
\end{equation}
with $u^{\phi} = \omega u^t$ and the other components of $u^{\mu}$ vanishing. The angular momentum of the shell as seen by an observer at infinity is $\omega$ and its surface density is $\sigma$ \cite{DeLaCruz}. We will require that $\omega$ and $\sigma$ are constant.

Our strategy for determining the metric induced by the shell is to compute the metric inside and outside the shell, perturbatively in $\omega$, 
\begin{equation}
	U = \sum_{i} U_{(2i)} \omega^{2i},\quad
	K = \sum_{i} K_{(2i)} \omega^{2i},\quad
	W=\sum_{i} W_{(2i)} \omega^{2i},\quad
	A=\sum_{i} A_{(2i)} \omega^{2i}
\end{equation}
and match the solutions across the shell using the Israel junction conditions \cite{israel1967nuovo}. 

To zeroth order in $\omega$ the shell is not rotating and we recover the Schwarzschild exterior and a flat interior:
\begin{align}
	U_{(0)} &= \begin{cases}
		\ln\left( (r-r_s)/(r+r_s) \right) & r \geq R \\
		\ln\left( (R-r_s)/(R+r_s) \right) & r < R
		\end{cases},
	\quad
	K_{(0)} = \begin{cases}
		\ln\left( (r^2-r_s^2)/r^2 \right) & r \geq R \\
		\ln\left(  (R^2-r_s^2)/R^2 \right) & r < R
		\end{cases}\\
\intertext{and}
	W_{(0)} &= e^{K_{(0)}}r \sin\theta\,.
\end{align}

The equations of motion at order $\omega$ involve only $A_{(0)}$. Requiring that the exterior geometry is asymptotically flat, that the interior is non singular, and that the induced metric on the sphere is continuous we obtain
\begin{equation}
		A_{(0)} = \begin{cases}
			\frac{r^3 \lambda_0}{(r+r_s)^6}& r \geq R \\
			\frac{R^3 \lambda_0}{(R+r_s)^6} & r < R
		\end{cases},
\end{equation}
with $\lambda_0$ an undetermined integration constant. Enforcing \eqref{E:energymomentum} for a rigidly rotating sphere implies that
\begin{equation}
	\lambda_0 = \frac{4 (2 R-r_s) r_s (R+r_s)^5}{R^3 (3 R-r_s)}\,.
\end{equation}
So far we have reproduced the results of \cite{brill1966rotating}.

The same procedure can be carried out at order $\omega^2$. After matching the induced metric along the rotating sphere we find that
\begin{equation}
	U_{(2)} = \begin{cases} 
		u_a(r) + u_b(r) L_2(\cos\theta) & r \geq R \\
		\beta_8 + \beta_9 r^2 L_2(\cos\theta)  & r<R
		\end{cases}\,,
	\qquad
	K_{(2)} = \begin{cases} 
		k_a(r) + k_b(r) (\sin\theta)^2 & r \geq R \\
		\beta_5 + r^2 \beta_6 - 2 r^2 \beta_6 (\sin\theta)^2& r<R
		\end{cases}
\end{equation}
and
\begin{equation}
	W_{(2)} = \begin{cases}
		\frac{\beta_1}{r^3} \sin(3\theta) + \frac{\beta_0}{r} \sin\theta & r \leq R\\
		\beta_7 r \sin\theta + \beta_6 r^3 \frac{R^2-r_s^2}{3 R^2} \sin(3\theta) & r<R
		\end{cases}\,,
\end{equation}
where $L_2$ is a second order Legendre polynomial, the $u$'s and $k$'s are given by
\begin{align}
\begin{split}
	u_a =& \frac{192 r^2 \left(r_s ^2 {\beta_0}+{\beta_1}\right)-128 r_s ^2
   {\beta_1}}{192 r_s ^3 r (r^2-r_s^2 ) } -\frac{\lambda_0 ^2 \left(13 r_s ^2 r^3+3 r \left(r_s ^2+r^2+3 r_s 
   r\right)^2\right)}{192 r_s ^3 (r+r_s)^6} + {\beta_3} \log \left(\frac{r-r_s }{r+r_s}\right) \\
	u_b =& \frac{8 \beta _1}{3 r r_s (r^2-r_s^2)}+\frac{\lambda _0^2 r^3}{12 r_s
   \left(r_s+r\right){}^6}+\frac{2 \beta _2 r_s \left(r^2+r_s^2\right)}{r}+\beta _2
   \left(\frac{r_s^4}{r^2}+r^2+\frac{2 r_s^2}{3}\right) \log \left(\frac{r-r_s}{r+r_s}\right) \\
	k_a =& \frac{\beta_0}{r^2-r_s^2}+ \frac{3 \beta_1}{r^2(r^2-r_s^2)} \\
	k_b =& -\frac{2 \beta_1(4 r^2 - 3 r_s^2)}{r^2(r^2-r_s^2)^2} + \frac{r^4 \lambda_0^2}{2(r^2-r_s^2)^2(r+r_s)^4}+\frac{r^2 \beta_4}{(r^2-r_s^2)^2} \\
		&- \frac{4 r_s \beta_2 \left( 2 r r_s(r^4+r_s^4) + (r^2-r_s^2)^2 (r^2+r_s^2)\ln\left(\frac{r-r_s}{r+r_s}\right)\right)}{r(r^2-r_s^2)^2}
\end{split}
\end{align}
and the $\beta_i$'s are determined in terms of $\beta_0$ via the linear equations
\begin{align}
\notag
	\beta_9 =& \frac{u_b(R)}{R} &
	\beta_8 =& u_a(R) &
	\beta_7 =& \frac{\beta_0}{R^2} \\
	\beta_6 =& \frac{3 \beta_1}{R^4(R^2 - r_s^2)} &
\label{E:betas}
	\beta_5  =& k_a(R)+\frac{1}{2}k_b(R) &
	\beta_4 =& \frac{2 \beta_1}{\alpha^2} - 6 \alpha^2 \beta_3 - \frac{\lambda_0^2}{32 \alpha^3} \\
\notag
	\beta_3 =& \frac{\beta _0}{2 r_s^2}+\frac{\beta _1}{2 r_s^4}-\frac{\lambda _0^2}{128 r_s^4} &
	\beta_2 =& -\frac{3 \beta_3}{8\alpha^2} &
	\beta_1 =& -\frac{1}{6}R^2 (R^2 - r_s^2)k_b(R) \, .
\end{align}

The value of $\beta_0$ is determined by the requirement that the surface energy on the sphere (c.f., equation \eqref{E:energymomentum}) is uniform. We find
\begin{align}
\begin{split}
\label{E:beta0}
	0=&R \left(r_s+R\right) k_b'(R)-2 k_b(R) \left(R - r_s\right)+3 R \left(r_s+R\right)
   u_b'(r)-3u_b(r) \left(R-r_s\right) \\ +
   &\frac{3 \lambda  R
   \left(\left(r_s+R\right){}^6-\lambda  R^3\right)}{2 \left(R-r_s\right)
   \left(r_s+R\right){}^6}+\frac{8}{3} \beta _6 R^2 \left(r_s+R\right)-3 \beta _9
   R^2 \left(r_s+R\right)+\frac{8 \beta _1 \left(2 R^2-r_s^2\right)}{R^2
   \left(R-r_s\right){}^2 \left(r_s+R\right)} \, . \\
\end{split}
\end{align}
Equation \eqref{E:beta0} together with \eqref{E:betas} determine all the $\beta_i$'s in terms of $R$ and $r_s$.

\end{appendix}

\bibliographystyle{JHEP}
\bibliography{KAnom}

\end{document}